\documentclass[preprint,showpacs,preprintnumbers,amsmath,amssymb]{revtex4}
\usepackage{graphicx}% Include figure files
\usepackage{dcolumn}% Align table columns on decimal point
\usepackage{bm}% bold math

\usepackage{hyperref}
\begin{document}

%\preprint{APS/123-QED}

\title{Generation of ultrabright tunable polarization entanglement
without spatial, spectral, or temporal constraints}

\author{Marco Fiorentino}
\email{mfiore@mit.edu}
\author{Ga\'etan Messin}

\author{Christopher E.\ Kuklewicz}
\author{Franco N.\ C.\ Wong}
\author{Jeffrey H.\ Shapiro}
\affiliation{Research Laboratory of Electronics, Massachusetts
Institute of Technology, Cambridge, MA 02139}

\begin{abstract}
The need for spatial and spectral filtering in the generation of
polarization entanglement is eliminated by combining two
coherently-driven type-II spontaneous parametric downconverters.
The resulting ultrabright source emits photon pairs that are
polarization entangled over the entire spatial cone and spectrum
of emission.  We detect a flux of $\sim$12\,000
polarization-entangled pairs/s per mW of pump power at 90\%
quantum-interference visibility, and the source can be temperature
tuned for 5 nm around frequency degeneracy.  The output state is
actively controlled by precisely adjusting the relative phase of
the two coherent pumps.
\end{abstract}

\pacs{03.65.Ud, 03.67.Mn, 42.50.Dv, 42.65.Lm}

\maketitle

Polarization entanglement has been used to demonstrate a variety
of quantum effects from quantum teleportation \cite{telep} to
quantum cryptographic protocols \cite{crypto}. The quality of
polarization-entangled photons sources can be characterized by
their pair flux and the purity of the entangled state they
generate \cite{kwiat94,kwiat95,kwiat99,shih,Brilliant}.
%For
%example, the pair generation rate is essential in quantum
%communications as it limits the bit transmission rate, and the
%purity of the state plays an important role in determining the
%level of security of a cryptographic quantum channel.
For the existing sources the requirements of high flux and high
purity are often in conflict. Consider, for example, type-II
spontaneous parametric downconversion (SPDC) in a noncollinearly
phase-matched beta-barium borate (BBO) crystal. Here
\cite{kwiat95} spatial and spectral filtering are necessary to
eliminate non-entangled photons that would reduce the purity of
the output state.
%Recently
%a SPDC source using two type-I phase-matched BBO crystals oriented
%orthogonally has been demonstrated \cite{kwiat99}. This source
%produces very high visibility and higher pair flux that the single
%crystal counterpart. Nevertheless spatial and spectral filtering
%are still necessary to achieve high state purity in this system.
A source of polarization-entangled photons has been proposed
\cite{kwiat94} and demonstrated \cite{shih} in which the outputs
of two different SPDC crystals are combined interferometrically.
It was recognized that such a setup would generate entangled
photons independent of their wavelengths and angles of emission
\cite{kwiat94}.  The two-crystal interferometer, however, did not
show the promised high flux and high visibilities \cite{shih};
this was attributed to technical difficulties in the alignment.

Our group has investigated the use of a collinearly-propagating
geometry and long periodically-poled crystals to simplify
alignment and to increase the output flux in both type-I
\cite{elliott} and type-II SPDC \cite{chris}. In the case of
type-II SPDC in periodically poled potassium titanyl phosphate
(PPKTP) \cite{chris}, we have obtained \emph{post-selected}
polarization-entangled photons. However, spatial and spectral
filtering are still required to obtain a high-purity entangled
state and the post-selection process involves a 3-dB loss. In this
Letter we report on a robust implementation of the coherent
addition of two SPDC sources based on a single PPKTP crystal. Our
scheme fully exploits, for the first time to our knowledge, the
properties of interferometric combining of two coherent SPDC
sources \cite{kwiat94} to yield an ultrabright source of
polarization entanglement that has no spatial or spectral
constraints. Moreover, collinear operation allows us to control
the output state by locking the pump phase of the same
interferometer.  This setup produces $\sim$10 times more
polarization-entangled pairs/s per mW of pump than any other
continuous-wave (cw) source in the literature
\cite{kwiat99,Brilliant}.

\begin{figure}[b]
\centerline{\rotatebox{0}{\scalebox{.5}{\includegraphics{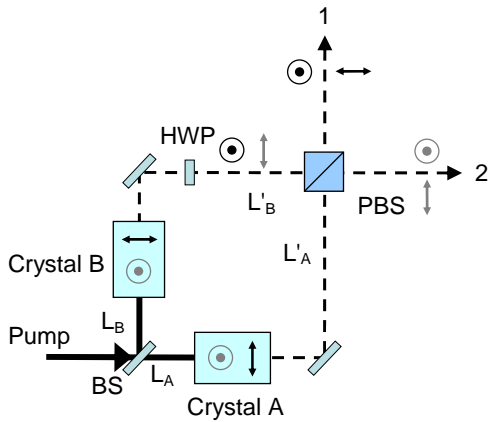}}}}
\caption{Schematic of the two-crystal source of
polarization-entangled photons.  Black (grey) refers to the signal
(idler) field amplitude at $\omega_s$ ($\omega_i$). Horizontal
(vertical) polarization: $\updownarrow$ ($\odot$). BS: beam
splitter. HWP: half-wave plate. PBS: polarizing beam splitter.}
\label{schematic}
\end{figure}

Figure \ref{schematic} illustrates a source that coherently
combines the outputs of two SPDC crystals. A laser is split by a
50--50 beam splitter (BS) and pumps the two crystals that are
phase matched for collinear type-II SPDC\@.  In the low-gain
regime, the bi-photon state just after the crystals is given by
\begin{equation}
\label{state1} \left| \Psi \right\rangle =
\frac{1}{\sqrt{2}}\left(\left|H_{A} (\omega_s) V_{A}( \omega_i)
\right\rangle + e^{i \phi_p} \left|H_{B} (\omega_s) V_{B}
(\omega_i) \right\rangle \right)\,,
\end{equation}
where $A$ and $B$ refer to the two arms of the interferometer,
$\omega_s$ and $\omega_i$ are the signal and idler frequencies,
respectively, and $\phi_p = k_p (L_B-L_A)$ is the difference of
the delays accumulated by the pump (with wave vector $k_p$) in the
paths $L_A$ and $L_B$ between the 50--50 BS and the crystals. A
half-wave plate (HWP) is used to rotate the polarizations by
90$^\circ$ in arm $B$, so that the output state after the
polarizing beam splitter (PBS) is
\begin{equation}
\label{state2} \left| \Psi \right\rangle = \frac{1}{\sqrt{2}}
\left( \left|H_{1} (\omega_s) V_{2} (\omega_i) \right\rangle +
e^{i \phi} \left|V_{1} (\omega_s) H_{2} (\omega_i) \right\rangle
\right)\,,
\end{equation}
where 1 and 2 refer to the two PBS output ports.  The overall
phase $\phi=\phi_p+ \phi_s +\phi_i$ is determined by the pump
phase $\phi_p$ and  the phase delays accumulated by the signal and
idler, respectively, with $\phi_{s,i} = k_{s,i} (L'_B-L'_A) -
\Delta \phi_{\lambda/2}(\omega_{s,i})$. The first term of
$\phi_{s,i}$ is the delay due to the arm lengths  $L'_A$ and
$L'_B$ between the crystals and the PBS, and the second term is
the phase difference introduced by the HWP\@.  Note that the phase
delays introduced by the identical crystals in the two arms
cancel.  Under collinear phase matching $k_p=k_s+k_i$, and $\phi$
is equal to the phase difference accumulated by the pump in the
Mach-Zehnder interferometer formed between the 50--50 BS and the
PBS except for a fixed offset due to the HWP\@.  The phase of the
output bi-photon state in Eq.\ \ref{state2} can therefore be
precisely controlled by locking the Mach-Zehnder interferometer as
seen by the pump alone: one can generate the triplet (for
$\phi=0$) or the singlet state ($\phi=\pi$), as well as
intermediate states that are linear combinations thereof.

\begin{figure}[b]
\centerline{\scalebox{.7}{\includegraphics{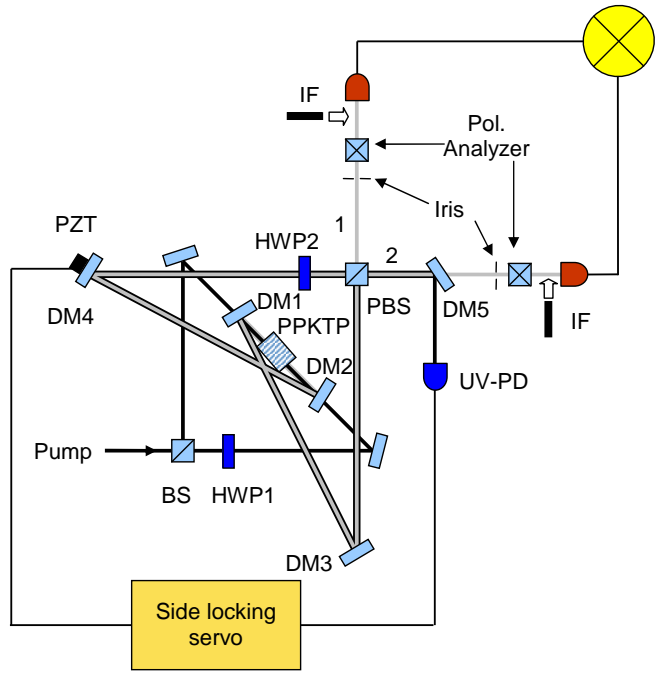}}}
\caption{Experimental setup. BS: 50--50 beam splitter. DM:
dichroic mirror. PBS: polarizing beam splitter. HWP: half-wave
plate. IF: 3-nm interference filter centered at 797 nm. HWP1 is
used to balance the flux of downconverted photons in the two
directions.} \label{setup}
\end{figure}

The HWP constrains the signal (idler) fields from the two crystals
to exit at output 1 (2) in Fig.~\ref{schematic}, ensuring
\cite{kwiat94,shapiro00} that the two sources are
indistinguishable so that all the photons are polarization
entangled regardless of their wavelengths and directions of
emission. Spatial and spectral filtering is unnecessary in this
two-crystal configuration, thus promising a source that has a much
higher photon-pair flux, plus a larger bandwidth and spatial
extension than BBO sources. Due to energy and momentum
conservation, one expects the emitted photon pairs from this
broadband spatially-extended source to show spectral and spatial
entanglement. Additional advantages of this scheme include
automatic erasure of timing distinguishability, non-degenerate
operation, and source tunability.

To implement the interferometric source described above it is
crucial that the two SPDC sources be identical. Source differences
introduce an element of distinguishability between the two paths
that would lead to a mixed state output. We therefore implemented
the scheme based on a single crystal with counter-propagating pump
beams derived from a single laser. The single-crystal approach is
particularly useful with periodically-poled crystals, as it
mitigates imperfections in the profiles of the periodic gratings.

We used a 10-mm-long ($X$ crystallographic axis), 1-mm-thick ($Z$
axis), and 4-mm-wide ($Y$ axis) hydrothermally-grown PPKTP crystal
with a grating period of 9.0 $\mu$m.  At a temperature of $\simeq
32^\circ$C this grating period phase matches type-II collinear
degenerate downconversion of a 398.5-nm pump polarized along the
$Y$ axis and propagating along the crystal's $X$ axis. The crystal
was housed in an oven and was maintained at its operating
temperature with $\pm$0.1$^{\circ}$C precision. This crystal was
previously characterized and used in type-II collinear SPDC to
yield single-beam quantum interference with a 99\% visibility
\cite{chris}.  We used second harmonic generation to measure the
temperature and wavelength tuning behavior in PPKTP using a cw
tunable laser centered around 797 nm. The second-harmonic
measurements are well described by the Sellmeier phase-matching
equations  for PPKTP \cite{sellmeier}, which allow us to calculate
the spatial and spectral properties of the downconverted photons,
as well as the phase-matching angles' dependence on the crystal
temperature. The latter predictions have been verified by imaging
the emitted photons with a CCD camera and narrow spectral filters.

The experimental setup is shown in Fig.\ \ref{setup}. The
frequency-doubled cw Ti:sapphire pump laser at 398.5 nm was split
by a BS that had a splitting ratio of $\sim$50--50. To balance the
powers of the two pump beams we inserted a half-wave plate (HWP1)
to vary the horizontally polarized pump power in the
(counter-clockwise propagating) brighter path.  The crystal was
not phase matched for a vertically polarized pump.  Each pump beam
focussed to a waist of $\sim$150 $\mu$m at the center of the PPKTP
crystal. The generated beams were collimated with 300-mm
radius-of-curvature dichroic mirrors (DM1,2) and combined at a PBS
after the polarization of one of the beams was rotated by
$90^\circ$ with a HWP\@.  The dichroic mirrors were coated for
high reflectivity (HR) at 797 nm and for high transmission (HT) at
398.5 nm, with a residual reflectivity of 0.2\% at the pump
wavelength. The pump beams, which propagated collinearly with the
downconverted beams, were weakly reflected by the four mirrors
(DM1-4) and recombined on the PBS, which had a $\sim$20--80
splitting ratio at the pump wavelength. The resultant pump beam
from port 2 of the PBS was directed  by a dichroic mirror (DM5, HR
at 398.5 nm and HT at 797 nm) for detection with an ultraviolet
photodiode (UV-PD). The BS and PBS in Fig.~\ref{setup} formed a
Mach-Zehnder interferometer for the pump and the detected fringes
were used to stabilize the interferometer with a side-locking
technique.  This provided a convenient and robust way to control
the phase $\phi$ of the output state in Eq.~2.  By inserting a
dispersive medium (such as a thin glass plate) in one of the arms
of the Mach-Zehnder interferometer we introduced a fixed but
variable offset between the phase of the pump fringes and the
phase of the output state (the overall offset phase includes other
dispersive elements in the interferometer).  By varying this phase
offset, we were able to lock the phase of the output state at an
arbitrary value while optimizing the side-locking feedback signal.

We placed two irises in the output beam paths to control the
acceptance angle of the detection system.  We estimate that an
iris diameter of 1 mm corresponded to an internal emission solid
angle of $\sim$3.5 $\times 10^{-5}$ sr at the crystal.  Flat
dichroic mirrors (not shown in Fig. \ref{setup}) similar to DM1
were used to eliminate residual pump light. The output photons
were detected with single-photon Si detectors (Perkin-Elmer
SPCM-AQR-14) through polarization analyzers (composed of a
half-wave plate and a polarizer).  The outputs of the
single-photon detectors were counted and also sent to an AND gate
(TTL logic family 74F) for coincidence counting. The coincidence
window for this configuration was measured to be $39.4$ ns. This
parameter allowed us to correct for the rate of accidental
coincidences in all of the data reported. For example, when
12\,000 coincidences/s were measured an average of 67\,000
singles/s were detected at each single-photon detector and $\sim
250$ coincidences/s were due to accidental Poisson processes.

\begin{figure}
\centerline{\rotatebox{270}{\scalebox{.35}{\includegraphics{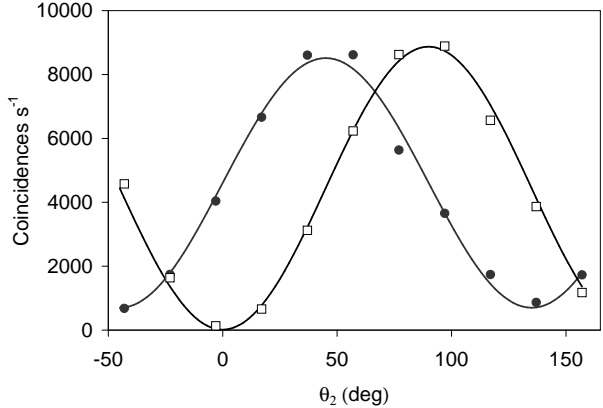}}}}

\caption{Coincidence counts for the frequency-degenerate singlet
state versus analyzer angle $\theta_2$ in arm 2 for analyzer angle
in arm 1 set at 45$^\circ$ (solid circles) and 0$^\circ$ (open
squares). Aperture size: 4-mm; no interference filter was used;
pump power: 0.7 mW\@. Each point is averaged over 10 s and the
lines are a sinusoidal fit to the data.} \label{results1}
\end{figure}

\begin{figure}
\centerline{\rotatebox{270}{\scalebox{.35}{\includegraphics{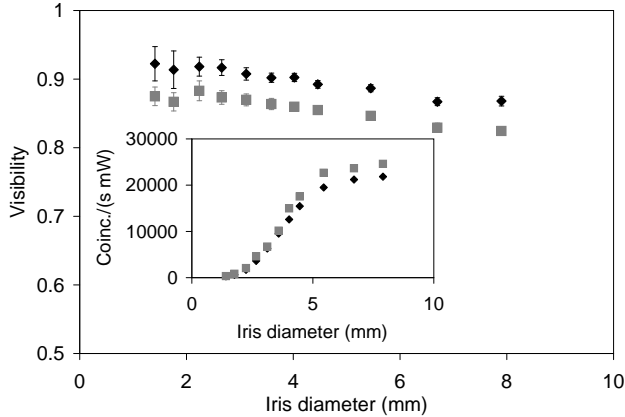}}}}
\caption{Frequency-degenerate singlet state 45$^\circ$-visibility
versus iris diameter. In the inset: coincidence counts/s per mW of
pump power versus iris diameter. No interference filter (squares)
and 3-nm interference filter centered at 797 nm (diamonds).}
\label{results2}
\end{figure}

\begin{figure}
\centerline{\rotatebox{270}{\scalebox{.35}
{\includegraphics{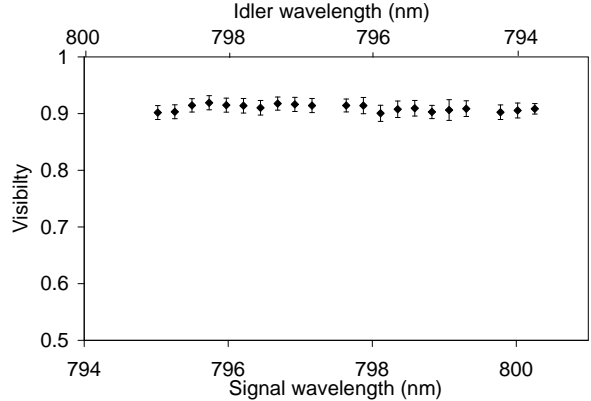}}}}
\caption{45$^\circ$-visibility (corrected for accidentals) versus
signal wavelength (no interference filter, iris diameter 2.2 mm)
with a measured flux of 3500 pairs s$^{-1}$ mW$^{-1}$.}
\label{results3}
\end{figure}

A summary of our experimental results, with the accidentals
removed, is shown in Figs.\ \ref{results1}--\ref{results3} for
$\phi=\pi$ (singlet). The temperature of the crystal was set to
$\sim$32$^\circ$C to ensure frequency degenerate operation. Figure
\ref{results1} shows the quantum interference in the coincidence
counts when the analyzer angle in arm 2 was varied for a fixed
angle in arm 1 with no narrowband interference filter. We observed
a visibility of 100 $\pm$ 3\% (85 $\pm$ 3\%) when analyzer 1 is
set to 0$^\circ$ (45$^\circ$). In what follows we will use the
45$^\circ$-visibility as an indication of the quality of the state
generated.

In Fig.\ \ref{results2} we report the 45$^\circ$-visibility for
the singlet state as a function of the iris diameter. Two sets of
data are shown, one using a 3-nm interference filters centered at
797 nm placed in front of the detectors (diamonds) and one in
which the interference filter was removed (squares). In both cases
the visibility is almost constant as a function of the iris
diameter. This allows us to increase the pair flux (Fig.\
\ref{results2} inset) while preserving the purity of the output
state. With the 3-nm filter we observed a visibility of 90\% and a
flux $\simeq$ 12\,000 pairs s$^{-1}$ mW$^{-1}$ with a 4-mm iris.
Under this condition, following Ref.\ \cite{Aspect}, we tested
Bell's inequality and obtained $S=2.599 \pm 0.004$, violating the
classical limit by more than 100 $\sigma$.

Figure \ref{results2} can be compared with data obtained in a
single-pass configuration with similar collection geometry
properties reported in Ref.\ \cite{chris}. The visibility of
quantum interference in the single-pass experiment drops much
faster as the iris diameter increases than in this interferometric
configuration.  The nearly constant visibility in
Fig.~\ref{results2} arises from effective spatial and spectral
indistinguishability in this dual-pumped interferometric
configuration.

Two main factors limited the visibility: wavefront distortion and
diffraction caused by the components of the interferometer, and
defects in the electric-field poling of the crystal.  Wavefront
distortion and diffraction lead to spatial distinguishability
between the two downconverted beams. Inhomogeneity in the crystal
grating introduces a temporal mismatch between the two paths. Both
these effects were mitigated somewhat by closing the iris and by
adding spectral filters.  To investigate the effects of wavefront
distortion and diffraction we measured the interferometer
visibility directly by injecting a laser beam at 797 nm through
arm 2 of the PBS in Fig.~\ref{setup} and observing the fringe
signal in arm 1. The input and output beam diameters could be
varied with irises.  When we changed the diameter of the output
beam for a fixed input beam diameter, the visibility showed the
same plateau for small iris diameters as in Fig.\ \ref{results2}.
When we decreased the input beam diameter for a fixed output beam
diameter the visibility increased linearly, approaching 100\%.
This suggests that the diffraction inside the interferometer was
responsible for the flat plateau in the visibility of Fig.\
\ref{results2}. We note that a slight mismatch in the length of
the two interferometric arms can also degrade the visibility.

No interference filter was used in obtaining the data shown in
Fig.\ \ref{results3} and the iris diameter was fixed at 2.2 mm.
The temperature of the crystal was then scanned between 20 and
50$^\circ$C and the 45$^\circ$-visibility was measured. We used
our knowledge of the Sellmeier equations to calculate the
phase-matched signal and idler wavelengths for each temperature
setting, and hence obtain the abscissas shown in this figure.
Figure \ref{results3} shows that the 45$^\circ$-visibility is
essentially independent of the signal and idler emission
wavelengths for a range of $\sim$5 nm around degeneracy.

In conclusion we have demonstrated a source of
polarization-entangled photon pairs with high flux and state
purity. The cw source is based on the interferometric combination
of two coherently-driven type-II sources of spontaneous parametric
downconversion from a single PPKTP crystal.   This dual-pumped
source is uniquely characterized by the fact that all the emitted
photon pairs are polarization entangled, regardless of their
wavelengths and directions of emission.  Therefore it can be
tuned, has a wide bandwidth and an extended spatial profile.  We
believe that our source produces spatial and spectral
entanglement, in addition to polarization entanglement, thus
providing additional degrees of freedom that can be used for
quantum communication. Further work with this source is needed to
experimentally demonstrate these additional forms of entanglement.
If successful, we would then have a source that could be used to
demonstrate fundamental quantum properties \cite{zeilinger} and in
cryptographic protocols with improved security \cite{genovese}.

This work was supported by a DoD Multidisciplinary University
Research Initiative (MURI) program administered by the Army
Research Office under grant DAAD-19-00-1-0177, the Defense
Advanced Research Projects Agency (DARPA) and Air Force Research
Laboratory under agreement F30602-01-2-0546, and the National
Reconnaissance Office.

\end{document}